\documentclass[aps,pre,twocolumn,showpacs,preprintnumbers,amsmath,amssymb]{revtex4-1}

\usepackage{graphicx}
\usepackage{dcolumn}
\usepackage{bm}
\setcitestyle{super}
\usepackage[latin1]{inputenc}
\usepackage{amsfonts}
\usepackage{hyperref}
\usepackage{amsmath}
\usepackage{amssymb}
\usepackage{amsthm}
\usepackage{bbold}
\usepackage{xcolor}
\usepackage{color}
\usepackage[american]{babel}
\usepackage[T1]{fontenc}
\usepackage{longtable}
\usepackage{xspace}
\usepackage{bbm}
\usepackage[all]{xy}

\newcolumntype{C}[1]{>{\centering\arraybackslash}p{#1}}

\begin{document}


\title{``On-the-fly'' calculation of the Vibrational Sum-frequency Generation Spectrum at the Air-water Interface}

\author{Deepak Ojha}
\affiliation{%
Dynamics of Condensed Matter and Center for Sustainable Systems Design, Chair of Theoretical Chemistry, Department of Chemistry, Paderborn University, Warburger Str. 100, 33098 Paderborn, Germany
}
\author{Thomas D. K\"uhne}
\email{tdkuehne@mail.upb.de}
\affiliation{%
Dynamics of Condensed Matter and Center for Sustainable Systems Design, Chair of Theoretical Chemistry, Department of Chemistry, Paderborn University, Warburger Str. 100, 33098 Paderborn, Germany
}
\affiliation{Paderborn Center for Parallel Computing and Institute for Lightweight Design,  Paderborn University, Warburger Str. 100, D-33098 Paderborn, Germany}

\date{\today}

\begin{abstract}
In the present work, we provide an electronic structure based method for the ``on-the-fly'' determination of vibrational sum frequency generation (v-SFG) spectra. The predictive power of this scheme is demonstrated at the air-water interface. While the instantaneous fluctuations in dipole moment are obtained using the maximally localized Wannier functions, the fluctuations in polarizability are approximated to be proportional to the second moment of Wannier functions. The spectrum henceforth obtained captures the signatures of hydrogen bond stretching, bending, as well as low-frequency librational modes.

\end{abstract}

\maketitle

\section{Introduction}

Vibrational spectroscopy provides microscopic fingerprints of the  structure and dynamics at the molecular level in condensed phase systems.\cite{tahara, shen1,shen2} However,  theoretical interpretation  and peak characterization of vibrational spectra  predominantly relies on molecular dynamics simulations.\cite{skinner,richmond,richmond2, tdk1,tdk2,spura} Nevertheless, the success of simulations also  depends largely on the forcefield employed to describe the interatomic interactions. In this regard, ab-initio molecular dynamics (AIMD) has proven to be  extremely useful as the interatomic forces are obtained from accurate electronic structure calculations.\cite{marx,CP2G}  For periodic systems, the overall electronic state within the AIMD framework is generally expressed in the terms of Bloch oribtals 
\begin{subequations}
\begin{equation} \label{Eq1a}
\Psi(r,k) = e^{(ik \cdot r)} \, u_{i}(r,k),
\end{equation}
where
\begin{equation} \label{Eq1b}
u_{i}(r,k) = u_{i}(r+R,k),
\end{equation}
\end{subequations}
with $\Psi(r,k)$ being the electronic wavefunction, $u_{i}(r,k)$ the Bloch function and $R$ a translational lattice parameter.\cite{ivo} An alternative representation, which is more suited for chemical problems, is provided by so-called maximally localized  Wannier functions (MLWFs), i.e.  $w_{n}(r-R)$ that are obtained by a unitary transformation of the Bloch orbitals.\cite{marzari,marzari2} The construction of this Wannier representation enables to split the continuously varying total electronic density into  contributions originating from localized fragments of the system. Mathematically, MLWFs are expressed as 
\begin{equation} \label{Eq2}
w_{n}(r-R) = \frac{V}{2\pi^3}\int_{BZ}d\mathbf{k} \, e^{-i\mathbf{k} \cdot R}\sum_{m=1}^{J}U_{mn}^{(k)}\psi_{m\mathbf{k}}(r), 
\end{equation}
where $R$ is the lattice vector of the unit cell and  V is the real-space primitive cell volume. The  $J \times J$ matrix $U_{mn}^{(k)}$  is the unitary transformation  matrix and $\psi_{m\mathbf{k}}(\mathbf{r})$ are the eigenstates of the system computed by density function theory (DFT). The corresponding MLWFs are then obtained by the unitary transformation $U_{mn}^{(k)}$ that minimizes the spread functional 
\begin{equation}\label{Eq3}
S = \sum_{n}S_{n} = \sum_{n}(\left \langle w_{n}\left | r^{2} \right |w_{n} \right  \rangle - \left \langle w_{n}\left | r \right |w_{n} \right \rangle^2).
\end{equation}
Therein, $\left \langle r^{2} \right \rangle $ is the second moment, whereas $\left \langle r \right \rangle^{2} $ is the squared first moment of the Wannier centers.
This unitary transformation based localization can be readily implemented on the position operator $\hat{r}$ within the Wannier representation to obtain localized orbitals for a given periodic system of arbitrary symmetry.\cite{silves,mpar,resta} As a result, the scheme can be used to compute the electronic contributions to the polarization of a system. Moreover, it also allows to  calculate instantaneous fluctuations in the molecular dipole moment and within the linear-response regime, obtain the linear as well as nonlinear infrared spectrum using time-correlation function formalism.\cite{galli,sprik,marx1,chao,ojha2,ac1,ac2} In this regard,  Raman and higher nonlinear analogs like v-SFG, 2D-vSFG and 2D-Raman can also be
computed  by applying a constant periodic electric field using the Berry phase formalism\cite{smith, resta2,umari}, or by calculating the  polarizability tensor $A$ \begin{equation}\label{Eq4}
A_{ij} = - \frac{\delta M_{i}(\mathbf{E})}{\delta E_{j}},
\end{equation}
where $M$ is the total dipole moment and $E$ is an externally applied electric field. This scheme of computing the polarizability tensor has been utilized to obtain isotropic Raman spectrum by means of density functional perturbation theory.\cite{baroni, putrino, luber, cp2k}

In this paper,  we present a novel  computational method  to obtain  the v-SFG spectrum of the air-water interface. This anisotropic Wannier Polarizability (WP) method is based on a technique of computing the fluctuations within the dipole moment and polarizbaility ``on-the-fly'' during an AIMD simulation without any additional computational cost.\cite{tdk3} For that purpose, the  fluctuations in the dipole moment are obtained using  the Wannier centers, whereas the components of the polarizability tensor are approximated using the second moment of the Wannier centers. However, it is noteworthy to mention that several other computational studies have obtained the vSFG spectrum using empirical maps\cite{morita1, morita2,morita3,skinner3,skinner4,skinner5,skinner6,skinner7}, velocity correlations\cite{yuki,yuki2,yuki3,naveen}, as well as directly from AIMD simulations\cite{cho,sulpizi,sulpizi2,ohto}.

The rest of the paper is structured as follows. In section 2, we outline our anisotropic WP method, whereas in section 3 the computational details to compute the v-SFG of the air-water interface are described. The corresponding numerical results are shown in section 4, before concluding remarks are provided in section 5.

\section{Anisotropic Wannier Polarizability  Method}

The original isotropic WP method has been implemented to compute the isotropic Raman spectrum of isolated gas-phase molecules, as well as aqueous solutions.\cite{tdk3,tdk4} The underlying principle of the method is that the polarization induced by an externally applied perturbation is directly proportional to the molecular volume of the system.\cite{bader1,bader2}  Since, Wannier centers provide a  picture, where the total electronic density is partitioned into the localized  electronic densities of different fragments of the system, the fluctuations in the electronic polarizability can be connected to the fluctuations of the volume of the  Wannier centers instead of the overall molecular volume.  As a result, the net isotropic polarizability can be expressed as 
\begin{equation}\label{Eq5}
\bar{A} = \frac{1}{3}\sum_{i=1}^{N_{WF}}A_{i}= \frac{\beta }{3}\sum_{i}^{N_{WF}}S_{i}^{3},
\end{equation}
where $S_{i}$ is the spread of the $i^{th}$ Wannier center, $N_{WF}$ is the number of MLWFs and $\beta$ is a proportionality constant. The isotropic Raman spectrum is then obtained as the Fourier transform of the polarizability time-correlation function.

On similar lines, the v-SFG spectrum of a non-centrosymmetric system is given as
\begin{equation} \label{Eq6}
\chi^{2}_{abc}(\omega) =  \int_{0}^{\infty }dt e^{i \omega t} \left \langle \dot{A}_{ab}(0) \cdot \dot{M}_{c}(t)  \right \rangle, 
\end{equation}
where $\chi^{2}_{abc}$ is the second order susceptibility, whereas $A_{ab}$ is the $ab^{th}$ component of the polarizability tensor  and $M_{c}$ is  $c^{th}$  component of the  dipole moment.\cite{morita1,skinner3,sulpizi} In contrast to Raman spectroscopy, the computation of v-SFG spectra requires the diagonal elements of the  polarizability tensor.  In this regard, we note that the second moment, i.e. $\left \langle w_{n}\left | r^{2} \right |w_{n} \right  \rangle $ and the polarizability are tensors of same size. Accordingly, we have approximated that the component specific  fluctuations in the polarizability are proportional to the second moment of the Wannier centers, i.e.
\begin{equation} \label{Eq7}
A_{ab} \propto \left \langle w_{n}\left | r_{ab}^{2} \right |w_{n} \right  \rangle.
\end{equation}
The strength of the anisotropic WP method is that for each set of an electron pair, we have a unique Wannier center and its corresponding moments. As a result, the method can be used to specifically study the contributions from the different fragments of the system. Moreover, it is also computationally less expensive as the polarizability is determined "on-the-fly" from the second moments of Wannier centers, which is in contrast with existing approaches, where the polarizability is obtained by numerical differentiation of the total dipole moment with respect to an externally applied electric field. This is to say that a simple minimization of the spread functional provides the Wannier centers and their corresponding moments, which are used to obtain the dipole and the polarizability, respectively. Thus, a single AIMD-based Wannier center calculation is sufficient to obtain the dipole moment, as well as the polarizability.

\section{Computational Details}
{ \em Ab initio} molecular dynamics simulations were performed by using the method of
Car and Parrinello,\cite{car,marx} as implemented in the  CPMD code\cite{cpmd}.
Simulations of the air-water interface comprising of 80 $H_{2}O$ molecules were performed at 300~K in a cubic box of edge length 12.43 \AA \ corresponding to the density at ambient conditions. \cite{lide} The air-water interface was generated by increasing the edge length of the box to 37.2 \AA \ in the z-direction.
The Kohn-Sham formulation of density functional theory was applied to represent 
the electronic structure  of the system within a plane wave basis set.\cite{CP2G} In order to represent
the  core-shell electrons, Vanderbilt ultra-soft pseudopotentials were
used and the  plane wave expansion of Kohn-Sham orbitals was truncated at a kinetic energy cutoff of 25 Ry.\cite{usp} The electronic orbitals were assigned a fictitious mass of 400 a.u. and equations of  motion were integrated with a time step of 4 a.u. 

In the present work, we have used the dispersion corrected BYLP-D exchange and correlation (XC) 
functional,\cite{becke,lee,grim2} since 
previous AIMD studies have shown that inclusion of London dispersion interactions not only improves the structure, but also predicts the dynamics, spectroscopy and phase diagram  of {\em "ab initio"} water and aqueous solutions in better agreement with experiment.\cite{ohto, mcgrath1, tdkAW, mcgrath2, yukiAW}
The initial configuration was generated using classical molecular dynamics simulations.
Subsequently, the production run was carried out in the canonical NVT ensemble using Nose-Hoover thermostats for 50~ps.

The identification of interfacial water molecules at the air-water system was conducted using the algorithm for the identification of truly interfacial molecules ITIM.\cite{itim,gitim}  This scheme uses a probe sphere to detect the molecules at the surface. The radius of the probe sphere was set to 2~\AA  \, which has been proven to be a good value for water.\cite{gitim}  A cutoff-based cluster search was also performed using 3.5  ~\AA  \ as a cut-off, which corresponds to the first minimum of the O$\cdots$O radial distribution function in liquid water. 

\section{Results and Discussion}

\begin{figure}
\begin{center}
\includegraphics[width=0.5\textwidth]{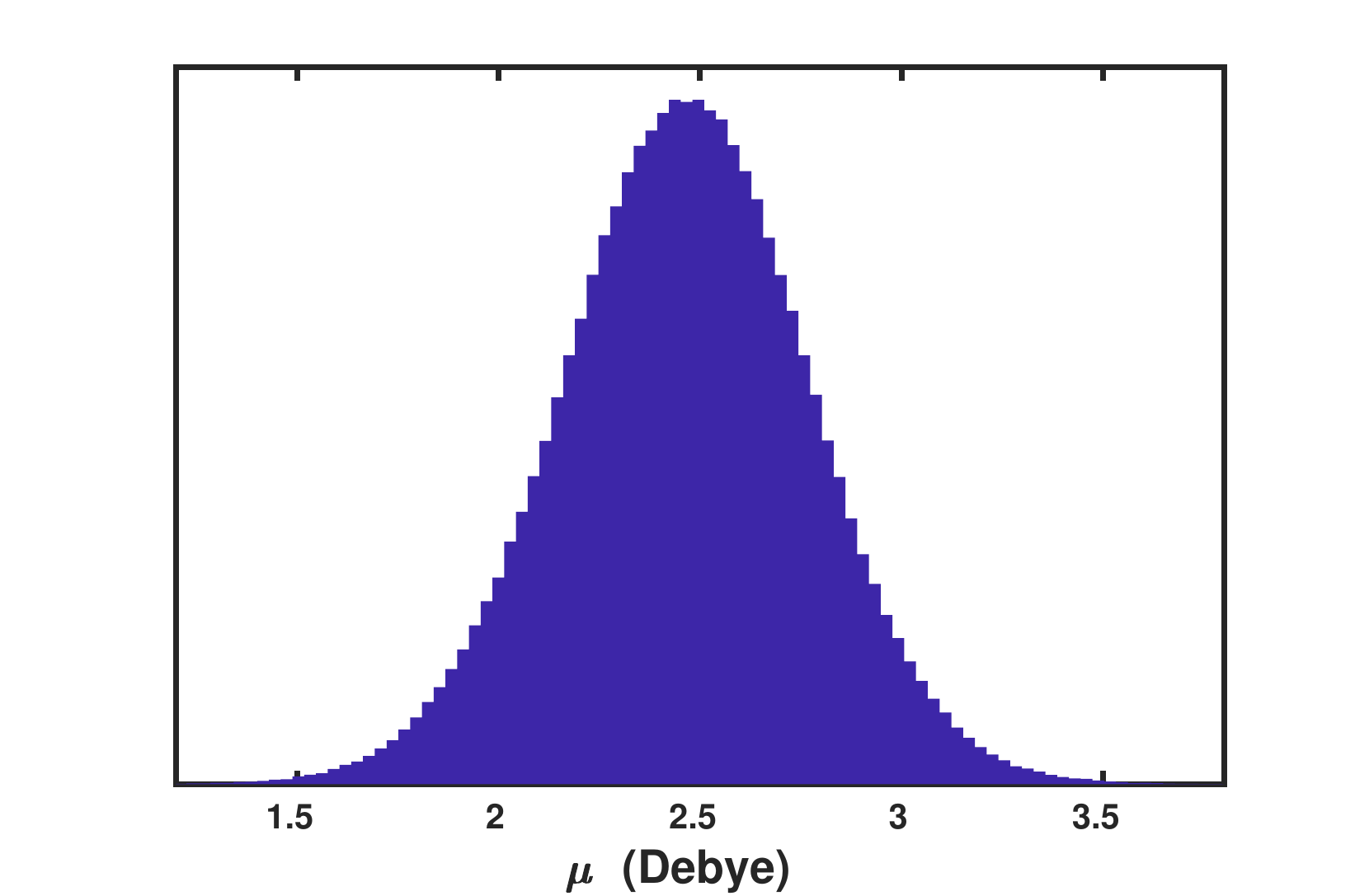}
\end{center}
\caption{\label{DipMom}
Distribution of molecular dipole moment ($\mu$) of water molecules at ambient conditions, as computed using the maximally localized Wannier centers.  
} 
\end{figure}
\begin{figure}[ht]
\centering
\includegraphics[width=0.5\textwidth]{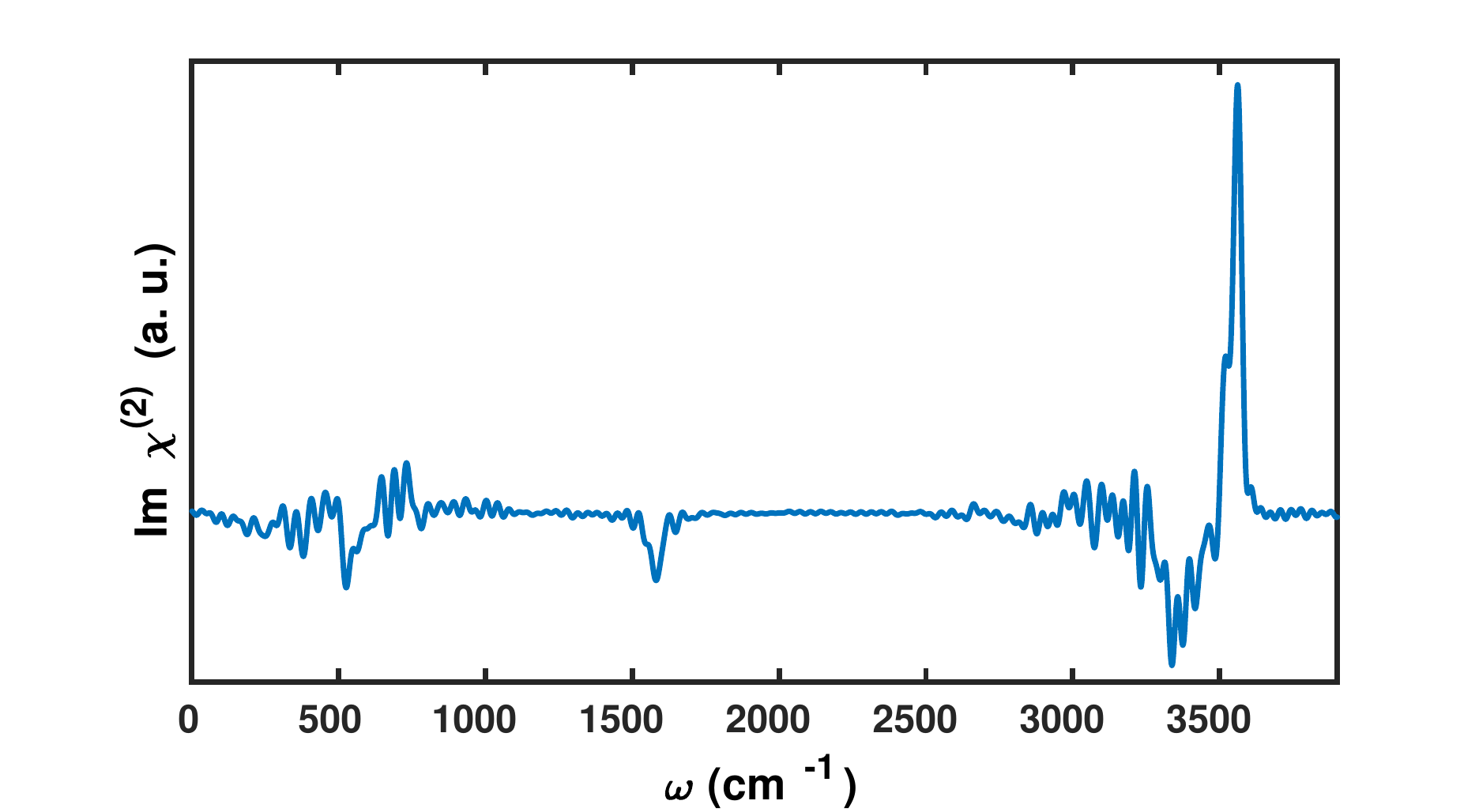}
\caption{\label{vSFG} The v-SFG spectrum of interfacial water molecules computed by the present anisotropic WP method.}
\end{figure}

To demonstrate the predictive power of the present anisotropic WP method, we have computed the v-SFG of at the air-water interface. For the sake of simplicity, we have assumed that the contributions originating from Wannier centers, which are associated with the lone pair of electrons, to the overall polarizability  can be safely neglected. The spectral dynamics is predominantly governed by the dynamical evolution of the Wannier centers corresponding to the bonded electron pairs. The average molecular dipole moment of the water molecules obtained using the Wannier centers, whose distribution is shown in Fig.~\ref{DipMom}, was found to be 2.46 Debye. 
\begin{figure}
\begin{center}
\includegraphics[width=0.5\textwidth]{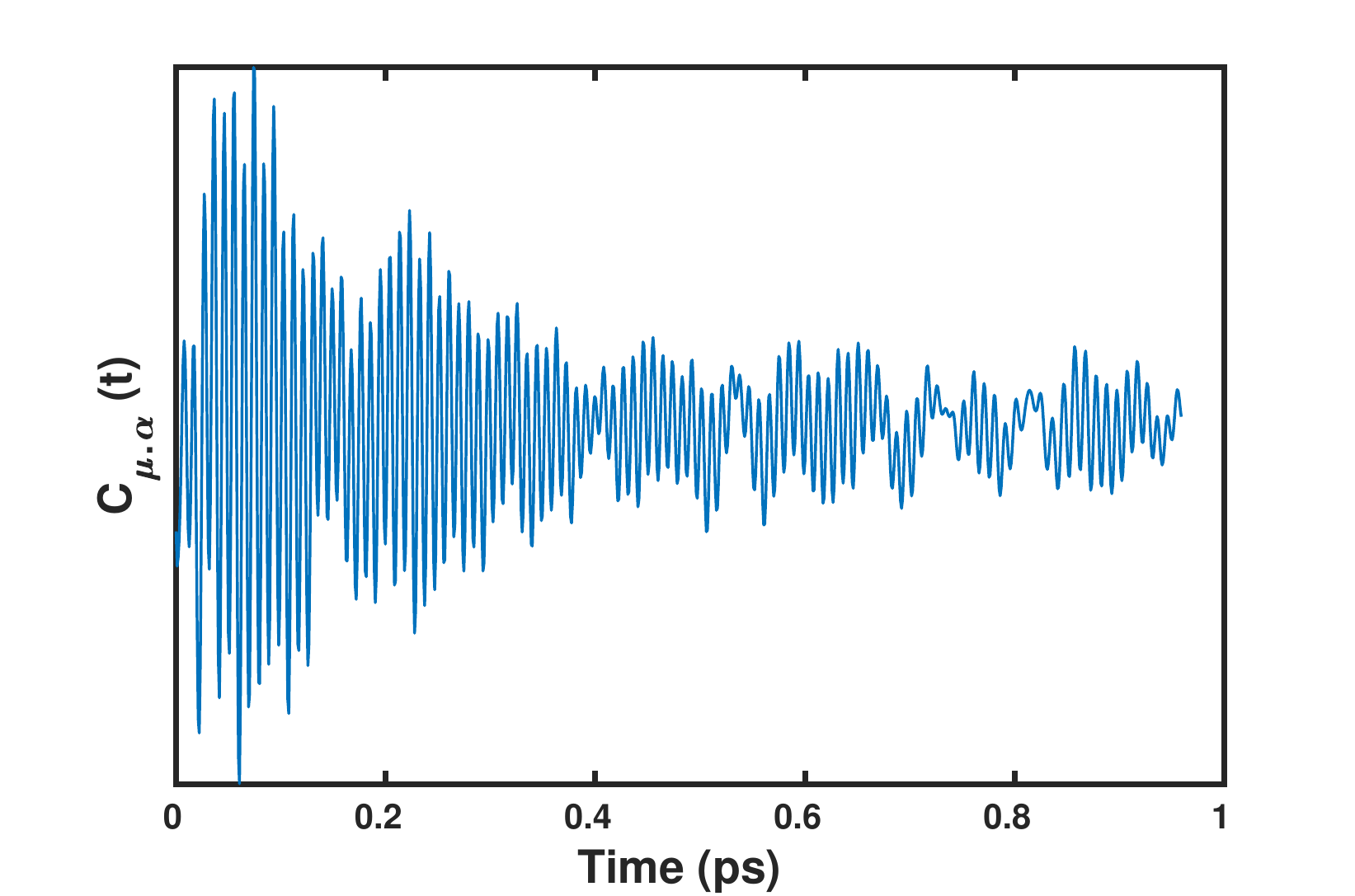}
\end{center}
\caption{\label{cross-correlation}
The dipole-polarizability cross-correlation function, as obtained by the present anisotropic WP method.  
} \end{figure}
The dipole-polarizability cross-correlation function and the v-SFG spectrum computed based on the fluctuations within the dipole moments obtained by using the Wannier centers and polarizabilities by means of the second moment are shown in Figs.~\ref{cross-correlation} and \ref{vSFG}, respectively.  
We find that the v-SFG spectrum obeys characteristic peaks corresponding to librational, bending, OH stretching, as well as free OH modes. Since there are various previous experimental and simulation based studies analysing the stretching, bending and librational modes within the v-SFG spectrum at the air-water interface, we only briefly highlight our findings in light of the existing literature. First, we will focus on the spectral region of 3000-3800 $cm^{-1}$, which is predominantly attributed to OH stretching modes. More precisely, earlier simulation studies have reported a broad negative peak between 3000 to 3600 $cm^{-1}$ and a sharp positive peak around 3700 $cm^{-1}$.\cite{morita1,morita2,skinner3,skinner4,sulpizi} Using our anisotropic WP method, we also find a broad negative peak at 2900-3500 $cm^{-1}$ and sharp positive peak around $\sim $ 3600 $cm^{-1}$. The former broad negative peak contribution originates from hydrogen-bonded water molecules with the overall dipole aligned towards the bulk, whereas the latter sharp positive peak is connected with the free and dangling OH modes of the interfacial water molecules.
The observed red-shift within the peak positions can be most likely attributed to the choice of XC functional and employed pseudopotentials in the present study. 
Earlier experimental and simulation studies of the bending mode have reported a broad negative peak around 1650 $cm^{-1}$ and a positive shoulder around 1750 $cm^{-1}$.\cite{yuki3,sulpizi2} Here, using the anisotropic WP method, we also observe a broad negative peak between 1400 and 1650 $cm^{-1}$, which is governed by the free and dangling OH modes. 
However, at variance to these earlier studies,\cite{yuki3,sulpizi2} we can not confirm any positive shoulder in our calculations. Finally, we observe a negative peak at around 450-650 $cm^{-1}$ that is governed by the librational motion of water molecules. 
Apart from a consistent red-shift within the peak positions, our results are in good agreement with earlier results that have also reported a negative peak in the region of 700-800 $cm^{-1}$.\cite{yuki2}

\section{Summary}
To summarize, we have proposed a computationally efficient ``on-the-fly'' method to determine the v-SFG spectrum for interfacial systems. This anisotropic WP method utilizes the second moment of the Wannier centers to estimate the polarizability fluctuations. The major strength of this method is  that it captures the spectral signatures of the system for the collective, as well as highly localized modes. Furthermore, it can be directly applied to spectral decomposition by computing fragment-specific contributions from the Wannier centers and their second moment to assist the interpretation of the experimental measurements. Moreover, the algorithm employed here can be easily extended to other spectroscopic techniques like two-dimensional v-SFG,\cite{mbonn} time-dependent v-SFG,\cite{ojha1} 2D-Raman-Thz,\cite{hamm} pump-probe Thz\cite{Mohsen} and 2D-Raman,\cite{mukamel} to name just a few. From the application perspective, interfacial reactivity, ``on-water'' catalysis, and other interfacial chemical processes can also be studied using our anisotropic WP-based method. Nevertheless, for greater agreement with the experiment, it would be important to better understand the role of simulation protocols and the approximations made, which we propose as an extensions for future works.

\begin{acknowledgments}
The authors would like to thank the Paderborn Center for Parallel Computing (PC$^2$) for the generous allocation of computing time on FPGA-based supercomputer ``Noctua''. This project has received funding from the European Research Council (ERC) under the European Union's Horizon 2020 research and innovation programme (grant agreement No 716142).
\end{acknowledgments}



\end{document}